\documentclass[aps,pof,reprint,10pt,superscriptaddress,showpacs]{revtex4-2}
\usepackage{amsfonts} 
\usepackage{amsmath}
\usepackage{amssymb}
\usepackage{graphicx}%\[  \]
\usepackage{subfigure}
\usepackage{color}
\usepackage{color}
\usepackage{soul}
\usepackage{cancel}
\usepackage{ulem}
\newcommand*\xbar[1]{%
  \hbox{%
    \vbox{%
      \hrule height 0.5pt % The actual bar
      \kern0.5ex%         % Distance between bar and symbol
      \hbox{%
        \kern-0.1em%      % Shortening on the left side
        \ensuremath{#1}%
        \kern-0.1em%      % Shortening on the right side
      }%
    }%
  }%
} 
\begin{document}
%\title{Influence of density inhomogeneity due to temperature variation on the convective instability of atmospheric stratified fluids}
%\title{Influence of density inhomogeneity on the convective instability due to thermal expansion of stratified fluids in the Earth's lower atmosphere }
\title{Influence of temperature-dependent density inhomogeneity on the  stability of atmospheric stratified fluids}
%\title{Influence of density inhomogeneity on the convective instability due to thermal expansion of atmospheric stratified fluids}
\author{T. D. Kaladze}
\email{tamaz{_}kaladze@yahoo.com}
\affiliation{I. Vekua Institute of Applied Mathematics and E. Andronikashvili Institute of Physics, Tbilisi State University, Georgia}
\author{A. P. Misra}
\homepage{Author to whom correspondence should be addressed}
\email{apmisra@visva-bharati.ac.in}
\affiliation{Department of Mathematics, Siksha Bhavana, Visva-Bharati University, Santiniketan-731 235, India}
\begin{abstract}
The stability of atmospheric stratified fluids is revisited to study the influence of the temperature-dependent density inhomogeneity due to thermal expansion in the Earth's lower atmosphere (with heights $0$ to $50$ km) under the action of gravity. Previous theory in the literature [Phys. Lett. A 480 (2023) 128990] is modified and advanced. It is found that the Brunt-V{\"a}is{\"a}l{\"a} frequency associated with internal gravity waves is modified, leading to new instability conditions of vertically stratified fluids. The possibility of the onset of  Rayleigh-B{\'e}nard convective instability is also discussed, and the influences of the modified Brunt-V{\"a}is{\"a}l{\"a} frequency and the density and temperature gradients on the instability growth rates are studied.

\end{abstract}
\maketitle
\section{Introduction} \label{sec-intro}
%Recently, Kaladze and Misra \cite{kaladze2023} studied the influences of the temperature and density gradients due to thermal expansion on the stability of vertically stratified fluids in the Earth's atmosphere. They showed that the instability due to thermal expansion can occur when the condition $\beta T_0>1$ is satisfied, where $\beta$ is the volumetric thermal expansion coefficient of incompressible fluids and $T_0$ is the background fluid temperature. In their approach, they ignored the dependence of the background density of stratified fluids on the variation of the background fluid temperature. However, spatial variations of fluid temperatures in the atmosphere can cause density variations owing to the thermal expansion \cite{acheson1973}. Thus, one can not ignore the dependence of the background density on the temperature variations. In this work, we consider this effect and generalize and modify the work by Kaladze and Misra \cite{kaladze2023}. We show that the Brunt-V{\"a}is{\"a}l{\"a} frequency associated with the internal gravity waves (IGWs) gets modified by this effect and obtain new instability conditions for IGWs. Consequently, the Rayleigh-B{\'e}nard convective instability, and hence the modified instability growth rate occurs.
%\par  
Continued global warming and climate change deserve great interest from scientists studying problems of atmospheric circulation dynamics, e.g., problems describing free thermal convection \cite{gershuni1977}. Such motions, which arise in the gravity field at spatial inhomogeneity of density caused by temperature inhomogeneity were intensively discussed.   Acheson and Hide \citep{acheson1973} noted that differential heating produces temperature variations from place to place, and owing to thermal expansion, they give rise to density stratification, causing so-called thermal instabilities. In \cite{tur2013a}, the effect of a small-scale helical driving force on fluid with a stable vertical temperature gradient, with a small Reynolds number, under the action of gravity is considered. The authors showed large-scale vortex instability in the fluid despite its stable stratification. At a nonlinear stage, this instability gives many stationary spiral vortex structures. Among these structures, there is a stationary helical soliton and a kink of the new type. Later in \cite{tur2013b}, authors considered stratified in gravity field rotating flow and found a new large-scale instability. They obtained nonlinear equations for the instability and gave a detailed study of the linear stage of the instability and the conditions of its appearance. They noted that instability is possible in the case of both stable and unstable vertical temperature stratifications. The authors in \cite{kopp2021} reported a new type of large-scale instability generated by the unstable vertical temperature gradient (heated from below) and small-scale force with zero helicity in an inclined rotating fluid with solutions in localized vortex kink-like structures.  Such investigations dealt with the propagation problems of internal gravity waves \cite{sutherland2010}. In Ref. \cite{yigit2015}, the authors discussed vertical coupling in the atmosphere and ionosphere system by internal waves generated in the lower atmosphere. In addition, they also discussed the progress in the sudden stratospheric warming and upper atmospheric circulation due to wave-induced vertical coupling between the lower and upper atmosphere.
\par   
Recently, Kaladze and Misra \cite{kaladze2023} studied the influences of the temperature and density gradients due to thermal expansion on the stability of vertically stratified fluids in the Earth's atmosphere.  The authors initiated an investigation in the atmospheric layer ($0< z < 50$ km heights) with negative and positive vertical temperature gradients in the tropospheric ($0< z < 15$ km)  and stratospheric $15< z < 50$ km) regions separately. They showed that such influence could lead to instability in stratified incompressible fluids. They also obtained the  Brunt-V{\"a}is{\"a}l{\"a} frequency modified by the thermal expansion coefficient and the critical value of the expansion coefficient for which the instability of internal gravity waves occurs.  In their approach, they ignored the dependence of the background density of stratified fluids on the variation of the background fluid temperature. However, spatial variations of fluid temperatures in the atmosphere can cause density variations owing to the thermal expansion \cite{acheson1973}. Such temperature dependence can significantly influence the spatio-temporal evolution of perturbed flows. Thus, it becomes necessary to take into account such effects in the dynamics of stratified atmospheric fluids. 
\par 
In this work, we consider such density inhomogeneity effect due to temperature variation and the vertical temperature gradient with an arbitrary sign, which allows the expansion of atmospheric layers with heights ranging from $0$ to $50$ km and to consider arbitrary length scales of density and temperature inhomogeneities. In this way, we modify and advance the work by Kaladze and Misra \cite{kaladze2023}. We show that the Brunt-V{\"a}is{\"a}l{\"a} frequency associated with the internal gravity waves (IGWs) gets modified by this effect, and we obtain new instability conditions for IGWs. Consequently, the Rayleigh-B{\'e}nard convective instability and the modified instability growth rate occur.
%%%%%%%%%%%%%%%%%%%%%
%In the given manuscript, we investigate the influence of temperature variation and atmospheric density inhomogeneity on the Earth’s atmosphere. The authors in \cite{kaladze2023} initiated such an investigation where the atmospheric layer ($15< z < 50$ km heights) with a positive vertical temperature gradient was considered. They showed that such influence could lead to instability in stratified incompressible fluids. They also obtained the  Brunt-V{\"a}is{\"a}l{\"a} frequency modified by the thermal expansion coefficient and the critical value of the expansion coefficient for which the instability of internal gravity waves occurs.  
%\par
%In the given paper, we systematize the theory carried out in [Kaladze, Misra] and consider the vertical temperature gradient with an arbitrary sign, which allows the expansion of atmospheric layers ($0< z < 50$ km) under the investigation.    
\section{Modeling of density inhomogeneity due to thermal expansion} \label{sec-model-density}
Typically, differential heating causes atmospheric convection in which the spatial variation of temperature may lead to the density variation due to the thermal expansion of stratified fluids. So, modeling of the density inhomogenity due to temperature variation is necessary for atmospheric fluid dynamics. To this end, we consider the following equation of state for incompressible atmospheric fluids, i.e., the dependence of the fluid density $(\rho)$ on the fluid temperature $(T)$ and pressure $(p)$.
\begin{equation}
\rho=\rho(T,p). \label{eq-rho1}
\end{equation} 
Any physical state variable, $f$ can be considered as the superposition of the mean (background) state and the perturbed state, i.e., $f=\overline{f}+f^\prime$, where the choice of the mean $\overline{f}$ plays a crucial role for the spatio-temporal evolution of the perturbation $f^\prime$.  
 Next, we consider the Taylor-series expansion of $\rho$ about the unperturbed or mean state  $\overline{\rho}(z)\equiv\overline{\rho}\left(\overline{T}(z),\overline{p}(z)\right)$ and assume that the deviations of the fluid density, pressure, and temperature from the unperturbed (background or mean) state are small. Thus, we obtain from Eq. \eqref{eq-rho1} the following expansion for $\rho$ to the first-order-smallness of perturbations.
\begin{equation}
\begin{split}
\rho({\bf r},t)&= \overline{\rho}(z)+\rho^\prime({\bf r},t)\\
&=\overline{\rho}(z)\left[1-\beta T^\prime({\bf r},t)+\alpha p^\prime({\bf r},t)\right], \label{eq-rho2}
\end{split}
\end{equation} 
where ${\bf r}=(x,y,z)$ and the expansion coefficients associated with the temperature and pressure, respectively, are
\begin{equation}
\beta=-\frac{1}{\overline{\rho}}\left(\frac{\partial \rho}{\partial T}\right)_{p},~
\alpha=\frac{1}{\overline{\rho}}\left(\frac{\partial \rho}{\partial p}\right)_{T}. \label{eq-bet-alp}
\end{equation}
The negative sign in $\beta$ is considered to mean that the fluid density increases with decreasing the temperature at atmospheric heights. The smallness of the perturbed density relative to its unperturbed state requires to fulfill the following condition:
\begin{equation}
|\beta T^\prime-\alpha p\prime|\ll1. \label{eq-cond-alp-bet0}
\end{equation}
Typically, the thermal expansion coefficient, $\beta$ is small $(\ll1)$, and it decreases with increasing the temperature \cite{kaladze2023}. So, the condition \eqref{eq-cond-alp-bet0} gives
\begin{equation}
|\beta T^\prime|\ll1,~~|\alpha p\prime|\ll1. \label{eq-cond-alp-bet1}
\end{equation}
Next, we assume that the fluid density variation due to the pressure inhomogeneity is small compared to the temperature inhomogeneity, i.e., 
\begin{equation}
|\alpha p^\prime|\ll|\beta T^\prime|. \label{eq-cond-alp-bet2}
\end{equation}
Such an assumption is valid for atmospheric stratified fluids. Thus, ignoring the term proportional to $\alpha$, Eq. \eqref{eq-rho2} reduces to
 \begin{equation}
\rho({\bf r},t)\equiv \overline{\rho}(z)+\rho^\prime({\bf r},t)=\overline{\rho}(z)\left[1-\beta T^\prime({\bf r},t)\right], \label{eq-rho3}
\end{equation}
where the perturbed density $\rho^\prime$ is given by
\begin{equation}
\rho^\prime({\bf r},t)=-\overline{\rho}(z)\beta T^\prime. \label{eq-rho-perturb}
\end{equation}
However, the density variation due to temperature can not be neglected as it gives rise to the fluid convective motion. The condition \eqref{eq-cond-alp-bet2} means that the pressure of the stratified fluids does not change drastically. In particular, it follows that the vertical scale of area where the convection occurs may not be too large. If the characteristic vertical scale is $L$, then the hydrostatic change of pressure has the order $\overline{\rho} gL$, i.e., $|p^\prime_1-p_2^\prime|\sim \overline{\rho}gL$, where $g$ is the constant acceleration due to gravity acting vertically downwards and the conditions \eqref{eq-cond-alp-bet1} and \eqref{eq-cond-alp-bet2} give  
\begin{equation}
\overline{\rho}gL|\alpha|\ll |\beta|\theta\ll1,\label{eq-cond}
\end{equation}
where $\theta=|T^\prime_1-T_2^\prime|$ is the characteristic temperature difference. 
\section{Basic Equations} \label{sec-basic}
 We consider the dynamics of incompressible stratified neutral fluids in the lower atmosphere (with heights  $0<z<50$ km) under the action of the gravity force ${\bf g}=(0,0,-g)$ per unit fluid mass density directed vertically downward. Our starting point is the following Navier-Stokes equation for incompressible neutral fluids. 
  \begin{equation}\label{eq-NS}
 \rho\left[\frac{\partial {\bf v}}{\partial t}+\left({\bf v}\cdot\nabla\right){\bf v}\right]=-\nabla p+\rho{\bf g}+\mu\nabla^2{\bf v}, 
\end{equation} 
where ${\bf v}$ is the neutral fluid velocity and  $\mu$ is the coefficient of the fluid dynamic viscosity.  The hydrostatic equilibrium state of fluid flow with ${\bf v}=0$ can be described from Eq. \eqref{eq-NS} as
\begin{equation}
0=-\frac{1}{\overline{\rho}}\nabla \overline{p}+{\bf g}. \label{eq-equbm-ns}
\end{equation}
%%%%%%%%%%%%%%%%%%%%%%%%%%%%
\par 
It is well known that the Boussinesq approximation is used to study convective motions or buoyancy-driven flows or gravity waves originating from the buoyancy force that are slower than typical sound waves in incompressible fluids. In this approximation, the kinematic viscosity coefficient is assumed to be constant, the density differences (due to inertial effects) are ignored except for the term associated with the gravity, and the density is solely a function of temperature. Thus, noting that $\rho$ behaves as the local fluid density and $\overline{\rho}$ as the reference density, and using the Boussinesq approximation and Eq. \eqref{eq-equbm-ns}, we obtain from Eq. \eqref{eq-NS} the following reduced equation for the perturbed fluid velocity ${\bf v}$ (without using prime symbol for ${\bf v}$, for simplicity).
 \begin{equation}\label{eq-NS2}
\frac{\partial {\bf v}}{\partial t}+\left({\bf v}\cdot\nabla\right){\bf v}=-\frac{1}{\overline{\rho}}\nabla p^\prime+\beta T^\prime g\hat{z}+\nu\nabla^2{\bf v}, 
\end{equation} 
where we have substituted $\rho^\prime=-\beta\overline{\rho} T^\prime$ from Eq. \eqref{eq-rho-perturb} and the term proportional to $\beta$ represents the buoyancy force. Also, $\nu=\mu/\overline{\rho}$ is the coefficient of the fluid kinematic viscosity and $\hat{z}$ is the unit vector along the positive $z$-axis (vertically upward). The momentum balance equation \eqref{eq-NS2} is to be closed by the following continuity equation and the heat equation in absence of any heat source for the perturbed velocity ${\bf v}$ and temperature $T^\prime$.
\begin{equation}\label{eq-cont}
\nabla\cdot{\bf v}=0,
\end{equation} 
\begin{equation}\label{eq-T}
 \frac{\partial T^\prime}{\partial t}+\left({\bf v}\cdot\nabla\right)(\overline{T}+T^\prime)=\chi\nabla^2 (\overline{T}+T^\prime),
\end{equation}
where  $\chi$ is the coefficient of the thermal diffusivity. In Eq. \eqref{eq-cont}, we have used the following condition for incompressible fluids.
\begin{equation}
\frac{d\rho}{dt}\equiv \frac{\partial \rho}{\partial t}+{\bf v}\cdot\nabla\rho=0. \label{eq-rho-const} 
\end{equation}
Next, from the basic state (unperturbed state) of heat equation \eqref{eq-T}, we have $d^2\overline{T}/dz^2=0$, which gives $d\overline{T}/dz=C_0$, a constant, and hence the following linear vertical temperature distribution.
\begin{equation}
\overline{T}(z)=C_0z+C_1,\label{eq-T1}
\end{equation}
where $C_1$ is another constant of integration. Applying this result to Eq. \eqref{eq-T}, we have 
\begin{equation}\label{eq-T2}
 \frac{\partial T^\prime}{\partial t}+\left({\bf v}\cdot\nabla\right)T^\prime+C_0w=\chi\nabla^2T^\prime,
\end{equation} 
where ${\bf v}\equiv(u,v,w)={\bf v}_{\perp}+w\hat{z}$. Furthermore, we recast Eq. \eqref{eq-rho-const} in the following reduced form.
\begin{equation}
w\frac{d\overline{\rho}}{dz}+\frac{d\rho^\prime}{dt}=0. \label{eq-rho-const1}
\end{equation}
Equations \eqref{eq-NS2}, \eqref{eq-cont}, \eqref{eq-T2}, and \eqref{eq-rho-const1}  are the desired set of equations for the dynamics of atmospheric stratified incompressible fluid flows. Comparing these equations with those in \cite{kaladze2023}, we note that not only the effect of fluid viscosity is considered in addition to the present work, the temperature (unperturbed)  gradient  $(C_0)$ appears to be constant in contrast to the work  \cite{kaladze2023}.   
\section{Brunt-V{\"a}is{\"a}l{\"a} frequency and instability condition} \label{sec-brunt}
To explicate the roles of the density gradient and the vertical temperature gradient, we consider a simple case, i.e., the linear flow of nonviscous fluids. Thus, separating the perpendicular  and vertical (along the $z$-axis) components of the fluid velocity in Eq. \eqref{eq-NS2}, we obtain 
\begin{equation}\label{eq-NS-perp}
\frac{\partial {\bf v}_\perp}{\partial t}+\frac{1}{\overline{\rho}}\nabla_\perp p^\prime=0,
\end{equation}
\begin{equation}\label{eq-NS-paral}
\frac{\partial w}{\partial t}+\frac{1}{\overline{\rho}}\frac{\partial p^\prime}{\partial z}-g\beta T^\prime=0.
\end{equation}
Also, the equation of continuity \eqref{eq-cont} gives
\begin{equation} \label{eq-cont1}
\nabla_\perp\cdot {\bf v}_\perp=-\frac{\partial w}{\partial z}.
\end{equation}
Taking divergence of Eq. \eqref{eq-NS-perp}, noting that $\nabla_\perp^2=\Delta_\perp=\partial^2/\partial x^2+  \partial^2/\partial y^2$, and using Eq. \eqref{eq-cont1}, we obtain
\begin{equation} \label{eq-w1}
\frac{\partial^2w}{\partial t\partial z}=\frac{1}{\overline{\rho}}\nabla^2_\perp p^\prime.
\end{equation} 
%%%%%%%%%%%%%%%%%%%%%%
Next, using the operator $\partial \Delta_\perp/\partial t$ to Eq. \eqref{eq-NS-paral} we get
\begin{equation}\label{eq-w2}
\frac{\partial^2}{\partial t^2}\Delta_\perp w+\frac{1}{\overline{\rho}}\frac{\partial^2}{\partial t\partial z}\Delta_\perp p^\prime-g\beta\frac{\partial}{\partial t}\Delta_\perp T^\prime=0.
\end{equation}
Also, using Eq. \eqref{eq-w1} and noting that $\overline{\rho}=\overline{\rho}(z)$, Eq. \eqref{eq-w2} reduces to 
\begin{equation} \label{eq-w3}
\frac{\partial^2}{\partial t^2}\left(\Delta w+\frac{1}{\overline{\rho}}\frac{d\overline{\rho}}{dz}\frac{\partial w}{\partial z}\right)-g\beta\frac{\partial}{\partial t}\Delta_\perp T^\prime=0,
\end{equation}
where  $\Delta=\Delta_\perp+\partial^2/\partial z^2$ is the Laplacian operator.
%%%%%%%%%
\par 
In what follows, we operate Eq. \eqref{eq-T2} with $\Delta_\perp$ to get in the linear approximation,
\begin{equation}\label{eq-T33}
\frac{\partial}{\partial t}\Delta_\perp T^\prime=\chi\Delta_\perp\Delta T^\prime-C_0\Delta_\perp w.
\end{equation}
Combining Eqs. \eqref{eq-w3} and \eqref{eq-T33}, we have
 \begin{equation} \label{eq-w4}
\frac{\partial^2}{\partial t^2}\left(\Delta w+\frac{1}{\overline{\rho}}\frac{d\overline{\rho}}{dz}\frac{\partial w}{\partial z}\right)-g\beta\chi\Delta_\perp\Delta T^\prime+gC_0\beta\Delta_\perp w=0.
\end{equation} 
Using the relation for the perturbed density, $\rho^\prime=-\overline{\rho}(z)\beta T^\prime$  [Eq. \eqref{eq-rho-perturb}]  to Eq. \eqref{eq-rho-const1}, we obtain
\begin{equation}\label{eq-cont2}
\left(1-\beta T^\prime\right)\frac{1}{\overline{\rho}}\frac{d\overline{\rho}}{d z}w-\beta\frac{dT^\prime}{dt}=0.
\end{equation}  
%%%%%%
Next, using Eq.\eqref{eq-T2} to Eq. \eqref{eq-cont2}, in the linear approximation, we obtain 
\begin{equation}\label{eq-cont33}
\frac{1}{\overline{\rho}}\frac{d\overline{\rho}}{d z}w -\beta\chi\Delta T^\prime+C_0\beta w=0.
\end{equation}
Finally, from Eqs. \eqref{eq-w4} and \eqref{eq-cont33}, we obtain the following evolution equation for stratified incompressible fluid flows in the lower atmosphere. 
\begin{equation}\label{eq-w5}
\frac{\partial^2}{\partial t^2}\left( \Delta w+\frac{1}{\overline{\rho}}\frac{d\overline{\rho}}{d z} \frac{\partial w}{\partial z}\right)+N^2(z)\Delta_\perp w=0,
\end{equation}
where $N^2$ is the squared Brunt-V{\"a}is{\"a}l{\"a} frequency, given by,
\begin{equation}\label{eq-N1}
N^2(z)=-\frac{1}{\overline{\rho}}\frac{d\overline{\rho}}{d z}g.
\end{equation}
Equation \eqref{eq-w5} has the same form as for ordinary internal gravity waves \citep{stenflo1987,kaladze2022} in absence of the effects of thermal expansion. 
It is important to note that the choice of the mean fluid density (background) is crucial for the spatio-temporal evolution of the perturbed fluid velocity $w$. Since the temperature variation in the Earth's atmosphere can cause the density variation due to thermal expansion, we model the background density $\overline{\rho}(z)\equiv \overline{\rho}\left(\overline{T}(z),\overline{p}(z)\right)$ as \cite{acheson1973}
\begin{equation}
\overline{\rho}(z)=\rho_0(z)\left[1-\beta \left(\overline{T}(z)-T_0\right)\right], \label{eq-rho0}
\end{equation}
where $\rho_0(z)$ is the fluid mass density at some reference temperature, $\overline{T}(z)=T_0$ at a particular height, applicable for ordinary internal gravity waves. In view of the assumption \eqref{eq-rho0}, the expression \eqref{eq-N1} for the Brunt-V{\"a}is{\"a}l{\"a} frequency  reduces to
\begin{equation}\label{eq-N2}
\begin{split}
N^2(z)=&\frac{g}{1-\beta {\cal T}}\times \\
&\left[\left(\beta {\cal T}-1\right)\frac{1}{\rho_0}\frac{d\rho_0}{d z} +\beta \frac{d\overline{T}}{d z} \right],\\
&=-g\left(\frac{1}{\rho_0}\frac{d\rho_0}{d z} -\frac{\beta {\cal T}}{1-\beta {\cal T}} \frac{1}{{\cal T}}\frac{d{\cal T}}{d z}\right),
\end{split}
\end{equation}
where $\rho_0\equiv \rho_0(z)$,  ${\cal T}\equiv{\cal T}(z)=\overline{T}(z)-T_0$.
Comparing the first form of Eq. \eqref{eq-N2} with Eq. (21) of Ref. \citep{kaladze2023}, we find that an additional factor $1-\beta {\cal T}(z)$ appears in Eq. \eqref{eq-N2} before the square brackets due to the modeling of the background density  with thermal expansion effects [\textit{cf}. Eq. \eqref{eq-rho0}]. Furthermore, in contrast to Ref. \citep{kaladze2023}, the temperature gradient $C_0\equiv d\overline{T}(z) /dz$ is no longer varying but a constant [\textit{cf}. Eq. \eqref{eq-T1}]. 
\par
\begin{figure*}
\centering
\includegraphics[width=6.5in, height=3in]{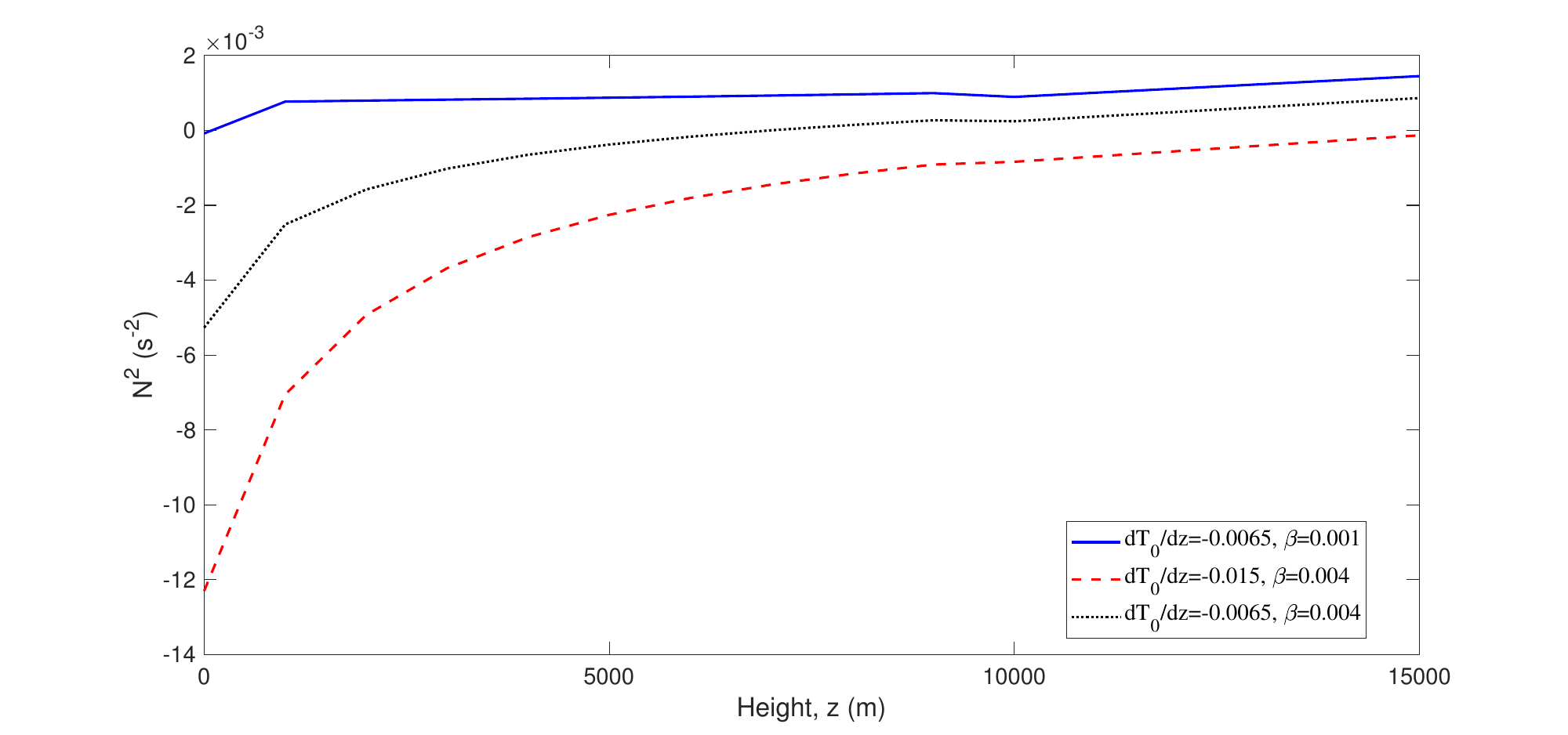}
\caption{ The squared Brunt-V{\"a}is{\"a}l{\"a} frequency ($N^2$) is plotted against the height ($z$) for different values of the thermal expansion coefficient $\beta$ as in the legend. The data used
are as in Table 1 of Ref. \cite{kaladze2023}. It is seen that for a fixed temperature gradient $dT_0/dz~(\mathrm{K/m})$, the instability occurs $(N^2<0)$ for $\beta\gtrsim0.004/\mathrm{K}$, otherwise the stratified fluid is said to be stable  $(N^2>0)$. }
\label{fig1N2}
\end{figure*}

 
In what follows, we obtain the instability conditions from Eq. \eqref{eq-N2}, and note that $d\rho_0/dz<0$ holds in the atmospheric heights $0<z<50$ km \cite{kaladze2023} and $\beta{\cal T}<1$ holds good as per the assumption \eqref{eq-cond}. Furthermore, ${d{\cal T}}/{d z}<0$ in the tropospheric heights  $0<z<15$ km and ${d{\cal T}}/{d z}>0$ in the stratospheric heights  $15<z<50$ km  \cite{kaladze2023}. So, in the tropospheric layer, the necessary conditions for the instability at which $N^2<0$  are $\beta{\cal T}<1$ and ${d{\cal T}}/{d z}<0$. On the other hand, since the conditions ${d{\cal T}}/{d z}>0$ and  $\beta{\cal T}<1$ are likely to hold in the stratosphere, the fluid is said to be stable there for $N^2>0$. The sufficient condition of instability gives 
\begin{equation}
\left(\frac{1}{\rho_0}\frac{d\rho_0}{d z} >\frac{\beta {\cal T}}{1-\beta {\cal T}} \frac{1}{{\cal T}}\frac{d{\cal T}}{d z}\right).\label{eq-cond-suff1}
\end{equation}
Since as said before, $d\rho_0/dz<0$  in the atmospheric region of heights $0<z<50$ km and  $\beta{\cal T}<1$ as per the assumption \eqref{eq-cond}, the condition \eqref{eq-cond-suff1} can only be satisfied when ${d{\cal T}}/{d z}<0$. Typically, the length scale of the temperature inhomogeneity is larger than that of the density inhomogeneity \cite{kaladze2023}, i.e.,
\begin{equation}
L^{-1}_{\cal T}\equiv\frac{1}{{\cal T}}\Big|\frac{d{\cal T}}{d z}\Big|<L^{-1}_{\rho_0}\equiv \frac{1}{\rho_0}\Big|\frac{d\rho_0}{d z}\Big|. \label{eq-cond-suff2}
\end{equation} 
So, by means of the relation \eqref{eq-cond-suff2},  Eq. \eqref{eq-cond-suff1} reduces to
\begin{equation}
1<\Big| \frac{L_{\cal T}}{L_{\rho_0}} \Big|<\frac{\beta {\cal T}}{1-\beta {\cal T}}. \label{eq-cond-suff3}
\end{equation} 
The condition \eqref{eq-cond-suff3} is satisfied only when $0.5\lesssim \beta {\cal T} \lesssim1$, which is rather a harsh one in the atmospheric regions. 
 Thus, we may conclude that incompressible stratified fluid flows may be unstable in the tropospheric layer $0<z<15$ km but stable in the other region, i.e., $15<z<50$ km. For typical values of the atmospheric fluid density $\rho_0(z)$ and temperature $\overline{T}$, and their gradients,  readers are referred to the work of Kaladze and Misra \cite{kaladze2023}. Using the data as in Table 1 of Ref. \cite{kaladze2023}, it is seen from Fig. \ref{fig1N2} that for a fixed temperature gradient, a critical value of the thermal expansion coefficient $\beta$ exists close to or above $(\beta\gtrsim0.004/\mathrm{K})$  which the instability of stratified fluid may occur in the tropospheric layer.  Such a critical value is in agreement with the previous work \cite{kaladze2023}. 
%%%%%%%%%%%%%%%%%%
\section{Rayleigh-B{\'e}nard convective instability}
As an illustration, we consider the possible convective instability in the tropospheric layer with negative vertical temperature gradient, i.e., ${d{\cal T}}/{d z}<0$. We assume that such layer spans between heights $z_0=0$ km and $z_1=15$ km. The temperatures at the bottom and top boundaries are kept fixed at $T=T_0$ at $z_0=0$ km and at $T=T_1$ at $z_1=15$ km, so that the temperature at the top boundary is lower than the bottom boundary, i.e., $\Delta T=T_0-T_1>0$. 
\par
As a starting point, we consider Eqs. \eqref{eq-NS2}, \eqref{eq-cont}, and Eq. \eqref{eq-T2}, and recast them in linearized  forms as  
\begin{equation}
\left(\frac{\partial}{\partial t}-\nu\nabla^2\right)u=-\frac{1}{\overline{\rho}}\frac{\partial p^\prime}{\partial x},\label{eq-u11}
\end{equation}
\begin{equation}
\left(\frac{\partial }{\partial t}-\nu\nabla^2\right)v=-\frac{1}{\overline{\rho}}\frac{\partial p^\prime}{\partial y},\label{eq-v11}
\end{equation}
\begin{equation}
\frac{\partial w}{\partial t}=-\frac{1}{\overline{\rho}}\frac{\partial p^\prime}{\partial z}+\beta g T^\prime+\nu\nabla^2w,\label{eq-w11}
\end{equation}
\begin{equation}
\frac{\partial u}{\partial x}+\frac{\partial v}{\partial y}+\frac{\partial w}{\partial z}=0,\label{eq-cont3}
\end{equation}
\begin{equation}
\frac{\partial T^\prime}{\partial t}-\Gamma w=\chi\nabla^2T^\prime, \label{eq-T3}
\end{equation}
where we have obtained the first three equations [\eqref{eq-u11} to \eqref{eq-w11}] from Eq. \eqref{eq-NS2}  after separating the velocity components along the axes and $\Gamma=-C_0=-d\overline{T}/dz>0$ is so called the lapse rate.
\par 
 Eliminating $T^\prime$ from Eqs. \eqref{eq-w11} and \eqref{eq-T3}, we get
\begin{equation}
\begin{split}
&\left[\left(\frac{\partial}{\partial t}-\chi\Delta\right)\left(\frac{\partial}{\partial t}-\nu\Delta\right)-\beta g\Gamma\right]w=\\
&-\left(\frac{\partial}{\partial t}-\chi\Delta\right)\frac{1}{\overline{\rho}}\frac{\partial p^\prime}{\partial z}. \label{eq-w44}
\end{split}
\end{equation}
Differentiating Eqs. \eqref{eq-u11} and \eqref{eq-v11} with respect to $x$ and $y$ successively and using the results for  $\partial u/\partial x$ and  $\partial v/\partial y$ in Eq. \eqref{eq-cont3}, we obtain
\begin{equation}
\left(\frac{\partial}{\partial t}-\nu\Delta\right)\frac{\partial w}{\partial z}=\frac{1}{\overline{\rho}}\Delta_\perp p^\prime. \label{eq-w55}
\end{equation}
Next, differentiating Eq. \eqref{eq-w44} with respect to $z$ we get
\begin{equation}
\begin{split}
&\left[\left(\frac{\partial}{\partial t}-\chi\Delta\right)\left(\frac{\partial}{\partial t}-\nu\Delta\right)-\beta g\Gamma\right]\frac{\partial w}{\partial z}=\\
&-\left(\frac{\partial}{\partial t}-\chi\Delta\right)\frac{\partial}{\partial z}\left(\frac{1}{\overline{\rho}}\frac{\partial p^\prime}{\partial z}\right). \label{eq-w44a}
\end{split}
\end{equation}
Further applying the operator $\partial/\partial t-\nu\nabla^2$ to Eq. \eqref{eq-w44a} and using Eq. \eqref{eq-w55}, we obtain
\begin{equation}
\begin{split}
&\left[\left(\frac{\partial}{\partial t}-\chi\Delta\right)\left(\frac{\partial}{\partial t}-\nu\Delta\right)-\beta g\Gamma\right]\frac{1}{\overline{\rho}}\Delta_\perp p^\prime=\\
&-\left(\frac{\partial}{\partial t}-\chi\Delta\right)\left(\frac{\partial}{\partial t}-\nu\Delta\right)\frac{\partial}{\partial z}\left(\frac{1}{\overline{\rho}}\frac{\partial p^\prime}{\partial z}\right), \label{eq-w44b}
\end{split}
\end{equation}  
or, 
\begin{equation}
\begin{split}
&\left[\left(\frac{\partial}{\partial t}-\chi\Delta\right)\left(\frac{\partial}{\partial t}-\nu\Delta\right)-\beta g\Gamma\right]\frac{1}{\overline{\rho}}\Delta_\perp p^\prime=\\
&-\left(\frac{\partial}{\partial t}-\chi\Delta\right)\left(\frac{\partial}{\partial t}-\nu\Delta\right)
\left(\frac{1}{\overline{\rho}}\frac{\partial^2 p^\prime}{\partial z^2}-\frac{1}{\overline{\rho}^2}\frac{d\overline{\rho}}{dz}\frac{\partial p^\prime}{\partial z}\right), \label{eq-w44c}
\end{split}
\end{equation} 
and finally we obtain the following equation.
\begin{equation}
\begin{split}
&\left(\frac{\partial}{\partial t}-\chi\Delta\right)\left(\frac{\partial}{\partial t}-\nu\Delta\right)\frac{\Delta p^\prime}{\overline{\rho}}-\beta g\Gamma\frac{1}{\overline{\rho}}\Delta_\perp p^\prime=\\
&\left(\frac{\partial}{\partial t}-\chi\Delta\right)\left(\frac{\partial}{\partial t}-\nu\Delta\right)\frac{1}{\overline{\rho}^2}\frac{d\overline{\rho}}{dz}\frac{\partial p^\prime}{\partial z}. \label{eq-w66}
\end{split}
\end{equation}  
We assume that the charateristic size of variation of $\overline{\rho}(z)$ is much larger than that for $p^\prime$. In this situation, $\overline{\rho}$ may be considered as a constant and we can recast Eq. \eqref{eq-w66} as
\begin{equation}
\begin{split}
&\left(\frac{\partial}{\partial t}-\chi\Delta\right)\left(\frac{\partial}{\partial t}-\nu\Delta\right)\Delta p^\prime-\beta g\Gamma \Delta_\perp p^\prime=\\
&-\frac{1}{g}\left(\frac{\partial}{\partial t}-\chi\Delta\right)\left(\frac{\partial}{\partial t}-\nu\Delta\right)N^2\frac{\partial p^\prime}{\partial z}, \label{eq-w77}
\end{split}
\end{equation} 
where $N^2$ is the Brunt-V{\"a}is{\"a}l{\"a} frequency given by Eq. \eqref{eq-N1}.
%%%%%%%
\par 
To reveal the role of the term proportional to $N^2$ on the wave dynamics, we assume that the length scale of fluid density inhomogeneity remains approximately a constant, i.e., $1/L_{\rho_0}\equiv \left(1/\rho_0\right)\left(d\rho_0/dz\right)$ is constant and so is $N^2$. The latter can assume a negative sign for the instability to occur (See Sec. \ref{sec-brunt}). Next, to study the characteristics of the wave eigenmode, we assume the perturbations to propagate as plane waves with wave vector ${\bf k}=(k_x,k_y,k_z)={\bf k}_\perp+k_z\hat{z}$ and wave frequency $\omega$ with constant amplitude $\tilde{p}$ in the following form:
\begin{equation}
p^\prime({\bf r},t)=\tilde{p}\exp\left(i{\bf k}\cdot{\bf r}-i\omega t\right). \label{eq-wave-form}
\end{equation}
Substitution of Eq. \eqref{eq-wave-form} into Eq. \eqref{eq-w77} yields the following dispersion relation for Rayleigh-B{\'e}nard convective flows.
\begin{equation}
\left(k^2-i\frac{N^2}{g}k_z\right)\left[\omega^2+i(\chi+\nu)k^2\omega-\chi\nu k^4\right]+\beta g\Gamma k^2_\perp=0. \label{eq-disp1}
\end{equation}
In particular, in absence of the effects of fluid viscosity and thermal diffusivity, i.e., $\nu=0,~\chi=0$, Eq. \eqref{eq-disp1} gives the following expressions for the real $(\omega_r=\Re\omega)$ and the imaginary $(\gamma=\Im\omega)$ parts of the wave frequency $\omega$.
\begin{equation}
\omega_r=\pm\frac{1}{\sqrt{2}}\frac{\sqrt{\beta g\Gamma}k_\perp}{K_N}\left(1-\frac{k^2}{K_N^2}\right)^{1/2}, \label{eq-omeg1}
\end{equation}
\begin{equation}
\gamma=\mp\frac{1}{\sqrt{2}}\sqrt{\frac{\beta\Gamma}{g}}N^2\frac{k_\perp k_z}{K_N^3}\left(1-\frac{k^2}{K_N^2}\right)^{-1/2}, \label{eq-gam1}
\end{equation}
where $\omega_r$ and $\gamma$ are related to
\begin{equation}
\omega_r\gamma=-\beta\Gamma N^2\frac{k_\perp^2 k_z}{K_N^4},
\end{equation}
and $K_N=\left(k^4+N^4k_z^2/g^2\right)^{1/4}$. From Eqs. \eqref{eq-omeg1} and \eqref{eq-gam1}, we note that for $N^2<0$ and $k_z>0$, we have the instability $(\gamma>0)$ and a real wave mode propagating vertically upwards in the atmosphere with the frequency $\omega_r~(>0)$.  Typically, for a particular atmospheric model, $|N^2|\sim10^{-4}-10^{-3}/\rm{s}^2$ and $g\sim10~\rm{m/s^2}$ \cite{kaladze2023} so that $K_N\approx k$ for $k_z\ll1$ such that $\omega_r\approx0$. However, for moderate values of $|N^2|$ and $k_z$, such an approximation may not be valid. Thus, in this approximation and  in absence of the effects of fluid viscosity and thermal diffusivity, the Rayleigh-B{\'e}nard convective flows with positive lapse rate $\Gamma$ can no longer propagate with a finite wave frequency $\omega_r$, but can have a finite growth rate of instability, given by,   
\begin{equation}
|\gamma|\approx\sqrt{\beta g\Gamma}\frac{k_\perp}{k}. \label{eq-gam2}
\end{equation}
Clearly, the instability grows and the instability growth rate increases with increasing values of the thermal coefficient $\beta$ and the thermal lapse rate $\Gamma$ without any cut-off unless $k_\perp=0$. Since the thermal lapse rate remains a constant, the increasing values of $\beta$ may correspond to the region of low-temperatures with heights ranging from $0$ to $15$ km (See, e.g., Table 1 of Ref. \cite{kaladze2023}).
\par 
On the other hand, if the dissipation due to thermal diffusion dominates over the time evolution of perturbation, the term $\omega^2$ can be neglected compared to the terms proportional to $\chi$. Thus, for a non-viscous $(\nu=0)$ slowly varying perturbations, Eq. \eqref{eq-disp1} gives, after separating the real and imaginary parts, the following wave frequency and the growth rate.
%\begin{equation}
%\omega_r=-\frac{\beta\Gamma N^2}{\chi K_N^4}\frac{k^2_\perp}{k^2}k_z\approx -\frac{\beta\Gamma N^2}{\chi}\frac{k^2_\perp}{k^6}k_z, \label{eq-omegar}
%\end{equation}
\begin{equation}
\omega_r=-\frac{\beta\Gamma N^2}{\chi K_N^4}\frac{k^2_\perp}{k^2}k_z, \label{eq-omegar}
\end{equation}
\begin{equation}
\gamma=\frac{\beta g\Gamma}{\chi K_N^4}{k^2_\perp}. \label{eq-gam2}
\end{equation}
%\begin{equation}
%\gamma=\frac{\beta g\Gamma}{\chi K_N^4}{k^2_\perp}\approx \frac{\beta g\Gamma}{\chi}\frac{k^2_\perp}{k^4}. \label{eq-gam2}
%\end{equation}
%%%%%%%%%%%%%%%%%% 
\begin{figure*}
\centering
\includegraphics[width=6.5in, height=3in]{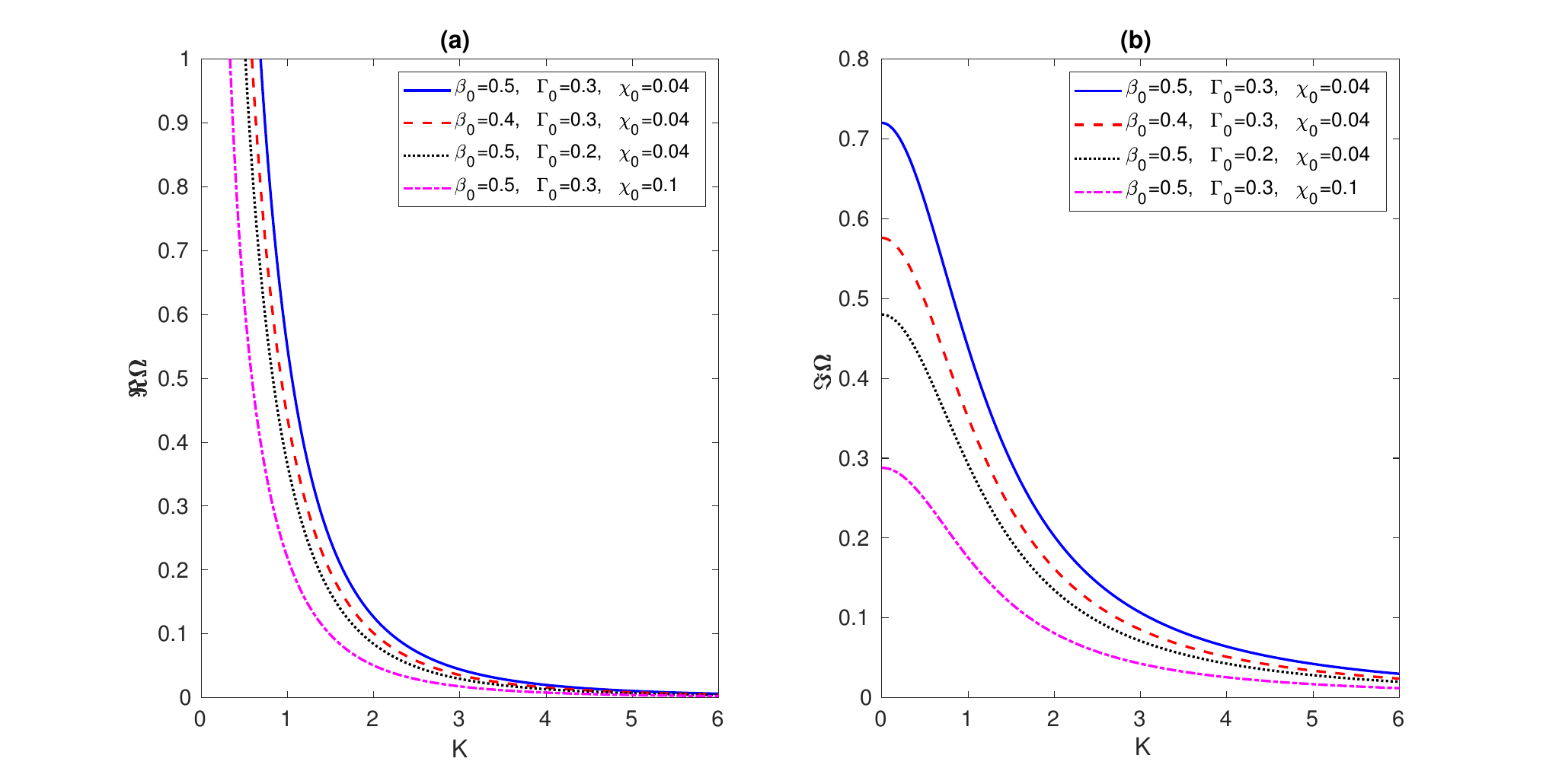}
\caption{The real [subplot (a)] and imaginary [subplot (b)] parts of the normalized wave frequency $(\Omega=\omega/\omega_g)$, given by Eqs. \eqref{eq-omegar} and \eqref{eq-gam2}, are plotted against the normalized wave number $(K=kL)$ for different values of the parameters as in the legends. Here, $N^2=-\omega_g^2<0$ and $L$ is the length scale of perturbations. The parameters are normalized as $\beta_0=\beta T_0$, $\Gamma_0=\Gamma L/T_0$, $\chi_0=\chi/\omega_g L^2$, and $\nu_0=\nu/\omega_g L^2$. Furthermore, $k_z=k\cos\theta$ and $k_\perp=k\sin\theta$ with $\theta$ denoting the angle between $k$ and $k_z$. The fixed parameter values are $\tilde{g}\equiv g/L\omega^2_g=0.4$, $\nu_0=0.05$, and $\theta=\pi/3$. The effect of $N^2$ is to reduce the instability growth rate (not shown). }
\label{fig1}
\end{figure*}
%%%%%%%%%%%%%%%%%%%%%%%% 
Since $\Gamma>0$ due to the assumption of the negative temperature gradient, we must have   $N^2<0$ for instability to occur (See Sec. \ref{sec-brunt}), and so  Eq. \eqref{eq-omegar} gives positive values of $\omega_r$.  From Eqs. \eqref{eq-omegar} and \eqref{eq-gam2}, it follows that both the wave frequency and the instability growth rate increase with increasing values of the thermal expansion coefficient, $\beta$ and the positive thermal lapse rate, $\Gamma$. In contrast, the thermal diffusivity effects significantly reduce the magnitudes of both $\omega_r$ and $\gamma$, since both vary inversely with $\chi$. Also, since the Brunt-V{\"a}is{\"a}l{\"a} frequency contributes to $\omega_r$ and $\gamma$ in the orders of $1/|N^2|$ and $1/N^4$ respectively, it can influence both the wave frequency and the growth rate in reducing their values at higher values of $|N^2|$. It is interesting to note that both the wave frequency and the growth rate tend to decrease with increasing values of the wave number $k$ and reach steady states at higher values of $k$. The characteristics of the real $(\omega_r=\Re\omega)$ and imaginary parts  $(\gamma=\Im\omega)$ are shown in Fig. \ref{fig1} for different values of the parameters $\beta$, $\Gamma$, and $\chi$. From the subplots, it is evident that for the propagation of gravity waves, the wave number $k$ should not be too small or too large for which the wave frequency exceeds the magnitude of the Brunt-V{\"a}is{\"a}l{\"a} frequency or it tends to vanish. Both the frequency and the instability growth rate attain maximum values at long-wavelength perturbations $(k\rightarrow0)$. 
%%%%
%Also, like Eq. \eqref{eq-gam2},  neither the sign nor the magnitude of $N^2$ can significantly influence the growth rate of instability. However, in contrast to it, the thermal diffusivity can influence the growth rate in increasing (decreasing) its values with decreasing (increasing) values of $\chi$.
%%% 
\par
\begin{figure*}
\centering
\includegraphics[width=6.5in, height=3in]{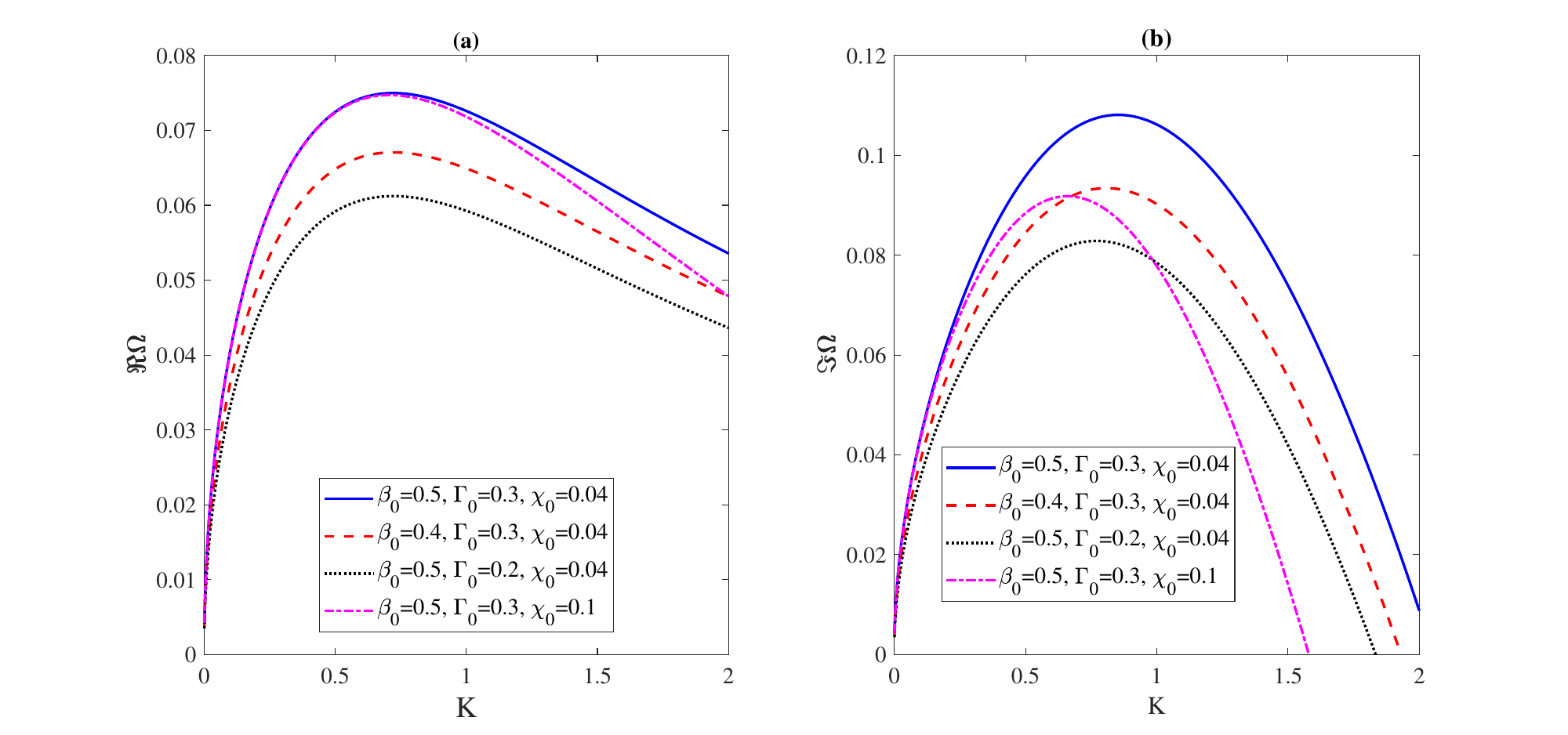}
\caption{The real [subplot (a)] and imaginary [subplot (b)] parts of the normalized wave frequency $(\Omega=\omega/\omega_g)$, given by Eq. \eqref{eq-disp1}, are plotted against the normalized wave number $(K=kL)$ for different values of the parameters as in the legends. Here, $N^2=-\omega_g^2<0$ and $L$ is the length scale of perturbations. The parameters are normalized as $\beta_0=\beta T_0$, $\Gamma_0=\Gamma L/T_0$, $\chi_0=\chi/\omega_g L^2$, and $\nu_0=\nu/\omega_g L^2$. Furthermore, $k_z=k\cos\theta$ and $k_\perp=k\sin\theta$ with $\theta$ denoting the angle between $k$ and $k_z$. The fixed parameter values are $\tilde{g}\equiv g/L\omega^2_g=1.5$, $\nu_0=0.05$, and $\theta=\pi/3$. The effect of $N^2$ is to reduce the instability growth rate (not shown). The effects of the viscosity $(\nu)$ on $\Re\omega$ and $\Im\omega$ are similar to $\chi$. }
\label{fig2}
\end{figure*}

Nevertheless, the features, stated before, can be a bit different in a more general situation in which the sign and magnitude of $N^2$ could influence the onset of instability and the growth rates rather than the possibility of damping due to viscosity and diffusion effects. So, we consider the general dispersion equation \eqref{eq-disp1} and solve it numerically. The results for the real and imaginary parts of the wave frequency are displayed in Fig. \ref{fig2}. We note that Eq. \eqref{eq-disp1} can be considered as a quadratic equation in $\omega$, which gives two branches of complex wave modes. However, we consider the branch that gives positive values of both the real and imaginary parts of the wave frequency. From the subplots of Fig. \ref{fig2}, we find that both the real wave frequency and the growth rate can achieve maximum values in domains of long-wavelength perturbations. Such maximum values can be further enhanced even with a small reduction of the values of $|N^2|$ (not shown in the figure). It is also noted that while the growth rate can have a cut-off, the real wave frequency approaches more or less a constant at higher values of the wave number. Furthermore, in both the cases, the effects of the thermal expansion $(\beta)$ and the temperature gradient $(\Gamma)$ are to enhance the wave frequency and the growth rate with their increasing  values (In agreement with the previous discussion at some  particular cases). Because of their appearance in the dispersion equation \eqref{eq-disp1}, any one of the coefficients of the fluid viscosity $(\nu)$ and the thermal diffusivity $(\chi)$ can influence both the wave frequency $(\Re\omega)$ and the instability growth rate $(\Im\omega)$. While their effects on $\Re\omega$ are not significant [See the dash-dotted line in subplot (a)], the growth rate can be significantly reduced with increasing values of any one of $\chi$ and $\nu$. For brevity, we have only shown graphically the effects of $\chi$ on the characteristics of $\Re\omega$ and $\Im\omega$.    
%%%%%%%%%%%%%%%%
%%%%%%%%%%%%%%%%%% 
%%%%%%%%%%%%%%%%%%%%%%%% 
\section{Discussion and conclusion}
We have investigated the influence of the temperature-dependent density inhomogeneity due to thermal expansion on the stability of stratified fluids in the Earth's lower atmosphere ($0<z<50$ km) owing to its importance in the spatio-temporal evolution of perturbed flows.  We show that modeling of such temperature dependence introduces a new factor involving the thermal expansion coefficient in the Brunt-V{\"a}is{\"a}l{\"a} frequency and thus modifies its expression and the instability condition and regions compared to those proposed in Ref. \cite{kaladze2023}. Furthermore, in contrast to Ref. \cite{kaladze2023}, the instability of stratified flows can occur for $\beta{\cal T}<1$ and $d\overline{T}/dz<0$ only in the tropospheric region $(0<z<15)$ km.  Consequently, the wave-eigenmode and the instability growth rate in the troposphere will be modified, and so will the evolution and the dynamics of internal gravity waves, governed by Eq. \eqref{eq-w5}, in the vertical stratified fluids. The detailed discussion along these lines is beyond the scope of the present investigation but a project of future work.  In the stratospheric region $(15<z<50)$ km, the fluid flow can be stable. It is also noted that the background temperature gradient, previously considered to vary in Ref. \cite{kaladze2023}, does not vary with atmospheric heights but remains a constant according to our assumption.
\par 
We also consider the influence of the modified Brunt-V{\"a}is{\"a}l{\"a} frequency, $N^2$ on the Rayleigh-B{\'e}nard convective flows and show that the negative temperature gradient for which the stratified flows associated with internal gravity waves become unstable can cause convective instability with an enhanced growth rate at a lower value of $N^2$. Such growth rates can be higher with increasing values of the thermal expansion coefficient and the thermal lapse rate, but can be significantly reduced by the effects of the thermal diffusion.
\par 
%%%%%%%%%%%%%%%%%
Our basic model has some limitations. For example, in the present investigation, we have neglected the fluid density variation due to the pressure inhomogeneity compared to the temperature inhomogeneity and the Earth’s rotation. Inclusion of these effects could again modify the  Brunt-V{\"a}is{\"a}l{\"a} frequency, and so is the instability condition of stratified fluids.  Furthermore, our modeling of the temperature-dependent density inhomogeneity and the absence of any heat source give rise to the constant gradient of the unperturbed temperature. However, the data in Ref. \cite{kaladze2023} suggests that the temperature gradient may not be precisely constant but may vary slightly with the Atmospheric height. We have also limited our discussion on the Rayleigh-B{\'e}nard convective instability to unbounded regions without estimating a critical Raleigh number beyond which the instability may occur. However, since the temperature can change sharply with the atmospheric height, the tropospheric region may be considered to be bounded between two heights with two different temperatures. Consideration of such bounded region together with the pressure inhomogeneity in the solution of Eq. (47) could significantly change the critical Raleigh number and the Nusselt number for critical modes and hence the instability growth rates.  We have also noted the possibility of thermal waves having a real-wave frequency and an instability-growth rate. Thus, one can calculate the vertical average heat flux for such unstable thermal waves to estimate the Nusselt number for critical modes.  Such unstable thermal waves at the nonlinear stage can open up several challenging areas to explore.  However, further investigations in these lines are underway, and we will communicate them elsewhere. Furthermore, since we have limited our discussion, particularly to the modification of the  Brunt-V{\"a}is{\"a}l{\"a} frequency and hence the instability condition of stratified fluids, as well as the Rayleigh-B{\'e}nard convective instability of linear wave modes, the direct numerical simulation of Eqs. \eqref{eq-NS2}, \eqref{eq-cont}, \eqref{eq-T2}, and \eqref{eq-rho-const1} have not been carried out but may be necessary for nonlinear evolution in our future studies.   
%%%%%%%%%%%%%%%%%
\par
The ongoing rise in global temperature significantly affects the Earth’s climate system, and its change can affect several atmospheric variables and the dynamics of atmospheric waves, e.g., internal gravity waves associated with them. The jump in temperature due to the absorption and release of heat by greenhouse gases causes the lower atmosphere (troposphere) to expand. Thus, predicting the instability of vertically stratified fluids due to thermal expansion and temperature gradients in the troposphere reported here could be crucial for realizing large-scale instability (may be larger than the scale of vortex or turbulence phenomena) through which the particle momentum and energy transfer can occur.   
%%%%%%%%%%%%%%%%%
\par 
As stated before, the proper modeling of the density inhomogeneity due to thermal expansion modifies the previously known instability condition \cite{kaladze2023}, i.e., the instability of stratified fluids may not occur in the entire region $0<z< 50$ km but more specifically in the tropospheric region:  $0<z< 15$ km. Thus, the present theory may more precisely predict the consequences of thermal instability (e.g., the nonlinear evolution and the dynamics of internal gravity waves) due to climate changes in the tropospheric region than that presented in Ref. 8.    
%%%%%%%%%%%%%%%%%%%
\par
To conclude, the density inhomogeneity-driven thermal instability of stratified fluids relating to the  Brunt-V{\"a}is{\"a}l{\"a} frequency could be helpful for the initiation of large-scale (may be comparable to the turbulence or vortex phenomena) instabilities through the momentum and energy transfer such as those associated with internal gravity waves and acoustic-gravity waves in the atmosphere \cite{rogachevskii2024}. Furthermore, due to the Rayleigh-B{\'e}nard convective instability, a more efficient energy transfer between two layers in the atmosphere can take place by the effects of the thermal expansion and the temperature gradient, leading to a transition from laminar to turbulent flows \cite{zhao2024,sharifi2024}. However, such efficient energy transport may eventually tend to resist some external perturbations, and the fluid flows may return to the equilibrium state. However, the detailed discussion concerning it is beyond the scope of the present study. 
\section*{Acknowledgments}
The authors thank the anonymous Referees for their insightful comments, which improved the manuscript in its present form.

%%%%%%%%%%%%%%%%%%%%%%%
%\section*{Acknowledgments} The authors thank all the three Referees for their insightful comments, which improved the manuscript in its present form.
%%%%%%%%%%%%%%%%%%%%%%%%%%%%% 
\section*{Author declarations}
\subsection*{Conflict of Interest}
The authors have no conflicts to disclose.
\subsection*{Author Contributions}
\textbf{T. D. Kaladze:} Writing--original draft, Validation, Methodology, Investigation, Formal analysis,  Conceptualization. \textbf{A. P. Misra:} Writing--review \& editing, Validation, Software, Methodology, Investigation, Conceptualization. 
%%%%%%%%%%%%%%%%%%%%%%%%%%
\section*{Data availability statement}
%The data that support the findings of this study are available from the corresponding author upon reasonable request.
All data that support the findings of this study are included within the article (and any supplementary files).
%%%%%%%%%%%%%%%%%%%%%%%%%5
\bibliographystyle{apsrev4-1} 
\bibliography{Reference}
\nopagebreak
\end{document}